\documentclass{aastex631}

\begin{document}

\title{Chemical Properties and Sagittarius-induced Dynamical Perturbations of the GD-1 Stream}
\shorttitle{GD-1 Stream}
\shortauthors{Liu et al.}

\correspondingauthor{Cuihua Du}
\email{ducuihua@ucas.ac.cn}

\author{Haoyang Liu}
\affiliation{School of Astronomy and Space Sciences, University of Chinese Academy of Sciences, Beijing 100049, P.R. China}

\author{Cuihua Du}
\affiliation{School of Astronomy and Space Sciences, University of Chinese Academy of Sciences, Beijing 100049, P.R. China}



\begin{abstract}
In this study, we investigate the chemical properties of the GD-1 stream using cross-matched, data-driven elemental abundances. The results reveal no clear $\alpha$-knee in the [Mg/Fe]-[Fe/H] plane, and strong abundance consistency between the thin stream and cocoon, supporting a common origin. The absence of multiple-population signatures (e.g., C-N anti-correlation) suggests a low-mass progenitor. Using a test-particle simulation with the particle spray method and including perturbations from the Sagittarius (Sgr) dwarf galaxy, it shows that Sgr does not significantly heat the stream to form the cocoon, but modifies the intrinsic $\phi_2$ distribution, in agreement with observations. The trailing arm narrowly distributed across the width of the stream, while the leading arm is more diffuse, indicating that major fraction of cocoon stars are present towards the leading arm. Sgr also drags more stream particles moving toward the Galactic center, producing an excess at $V_{\text{GSR}}<0$, consistent with data. Our study confirms the Sgr has a non-negligible dynamical influence on the GD-1 stream. Other heating mechanisms (e.g., dark matter sub-halo encounters and pre-stripping process inside the parent halo) remain to be considered, and higher-resolution spectroscopy is needed to further constrain chemical abundances.
\end{abstract}

\keywords{Stellar streams (2166) --- Milky Way dynamics (1051) ---  Chemical abundances(224)  ---  Globular star clusters (656)}


\section{Introduction} \label{sec:intro}
The Milky Way (MW) contains a lot of substructures recording the assembly history of the Galaxy itself, including the tidally disrupted dwarf galaxies \citep[e.g.,][]{Naidu2020,Malhan2022pontus,Tenachi2022}, the ongoing accretion event of the Sagittarius (Sgr) dwarf galaxy \citep{Ibata1995,Majewski2003}, and stellar streams originating from globular clusters (GCs) system \citep[see the review:][]{Bonaca2025}.
Among those substructures, the dynamically cold, metal-poor and long (extending up to 100 degree) GD-1 stream is one of the most extensively studied stellar streams and is considered to be the tidal debris of a disrupted GC \citep{Grillmair2006,Webb2019}. The observed gaps and off-track features (known as ``spur" and ``blob") of the GD-1 are considered to be the result of close encounters with dark matter sub-halos \citep{Carlberg2013,Bonaca2019,deboer2020,Doke2022}, thus used as a valid probe for studying the dark matter nature \citep{Mestre2024,Zhang2025}, as well as the constraints for the MW gravitational potential and mass \citep{Koposov2010,Malhan2019MNRAS,Nibauer2025}. 

Beside these features, \citet{Malhan2019} discovered a diffuse component (dubbed as ``cocoon") around the thin stream with a span of $\sim100$ pc \citep[also see][]{Malhan_Kshir}. Based on cosmological simulations, they interpreted this cocoon as the result of tidal stripping of the progenitor globular cluster within the parent dwarf galaxy sub-halos \citep[also see the pre-stripping process in][]{Carlberg2020,Qian2022}. With expanded GD-1 member stars from the Dark Energy Spectroscopic Instrument \citep[DESI; ][]{Desi2022}, \citet{Valluri2025} find that the cocoon is kinematically hotter with a velocity dispersion $\sigma\sim5-8$ km/s compared to that of the thin stream ($\sigma\sim3$ km/s) and the two components share a common origin because of the similar metallicity distributions. Because the velocity dispersion of the cocoon is comparable to that of dwarf galaxies, the authors suggest that the cocoon may also originate from a dwarf galaxy, while the thin stream likely comes from tidal debris of a nuclear cluster or a GC. Additionally, they propose three other possible scenarios for the origin of the cocoon component, including the influence of the Sgr, which we primarily focus on in the subsequent research.

Based on high-resolution spectroscopy, \citet{Bonaca2020} showed that the flyby of a massive, compact object (which is likely a black hole) could alter stellar orbits within the stream, creating the observed ``spur" feature. Using test-particle simulations, \citet{Dillamore2022} found that the Sgr could significantly affect GD-1-like streams, folding and twisting their morphology. Their best-fit model to the GD-1 data also indicates that an initial Sgr mass of $\gtrsim4\times10^{10} M_{\odot}$ is more compatible with streams formed more than 3 Gyr ago. However, they did not discuss the potential impact of the Sagittarius dwarf galaxy on the diffuse cocoon component. Given that Sgr did have a pericentric passage near the GD-1 stream, it is crucial to conduct a study on its role in forming the cocoon stars, rather than just focusing on its morphological impact.

To better understand the origin of the GD-1 stream and cocoon, chemical tagging also plays an equivalent role. \citet{Balbinot} detected C and Mg abundance variations at a significance level of $3\sigma$, similar to those found in low-mass Galactic GCs. Until recently, \citet{Zhao2025} conducted detailed abundance analyses of seven stars in the GD-1 stream. They found no clear multiple-population (MP) signatures but did identify a tight correlation between Eu and Ba abundances, which also implies a low-mass nature for the progenitor of GD-1. However, their sample is too limited to fully characterize the properties of GD-1 member stars in a statistically robust manner. In this data-driven era, elemental abundances inferred through deep learning provide a new opportunity for us to understand the origins of these substructures \citep[e.g.,][]{Leung2019,Berni2025}. Although these inferred elemental abundances face the risk of underestimated uncertainties, they can still provide statistically meaningful analyses on enlarged samples \citep{Liu2025}. Therefore, using these deep-learning-derived elemental abundances to analyze the GD-1 stream is of particular significance and constitutes one of the main focuses of our study.

In this work, we aim to investigate the properties of GD-1 based on a large sample using elemental abundances inferred from deep learning, while also taking the influence of the Sgr into account to examine whether the Sgr could be a possible origin of the cocoon component. This paper is structured as follows: Section~\ref{data} describes the literature from which we obtain the GD-1 member stars and the used elemental abundances. Section~\ref{results} presents the chemical analysis, as well as the setup and results of the corresponding test-particle simulations. Section~\ref{summary} gives a comprehensive summary of the results.

\section{data} \label{data}
\subsection{Member Stars of the GD-1 Stream}
Most member stars of GD-1 come from the work of \citet{Valluri2025}. These member stars were first selected based on spatial location and parallax, followed by further selection using isochrones. Due to the coverage of DESI observations, stars with $|\delta \phi_2|>3^{\circ}$ were also filtered out. The detailed procedure can be found in Section 3 of \citet{Valluri2025}. To obtain a more complete sample of GD‑1, we also directly include the high-probability member stars identified by \citet{Malhan2019GD} and \citet{Huang2019}, as well as the GD‑1 member stars with probabilities greater than 0.8 selected by \citet{Starkman2025} using mixture density networks. In subsequent abundance matching, we adopt stars from other studies when they overlap with \citet{Valluri2025}, and include their stars only when no duplicates exist. In total, there are 176 unique GD-1 member stars.

\subsection{Elemental Abundances of the GD-1 Stream}
To date, aside from \citet{Balbinot} and \citet{Zhao2025}, there have been no studies on the GD‑1 stellar stream providing more detailed chemical information beyond metallicity. Therefore, in order to study the elemental properties of GD‑1 with a larger sample, we decided to use element abundances inferred from deep learning. Considering that most member stars come from DESI observations, we adopted the elemental abundances for DESI stars from \citet{Zhang2024}, inferred using the DD‑Payne method, which models stellar spectra with a two-layer perceptron neural network. Moreover, \citet{Liu2025} provided a relatively large sample of elemental abundances for over one million giant stars from LAMOST DR10, with uncertainties within acceptable limits.

\begin{figure*}
    \begin{center}
        \includegraphics[width=1\textwidth]{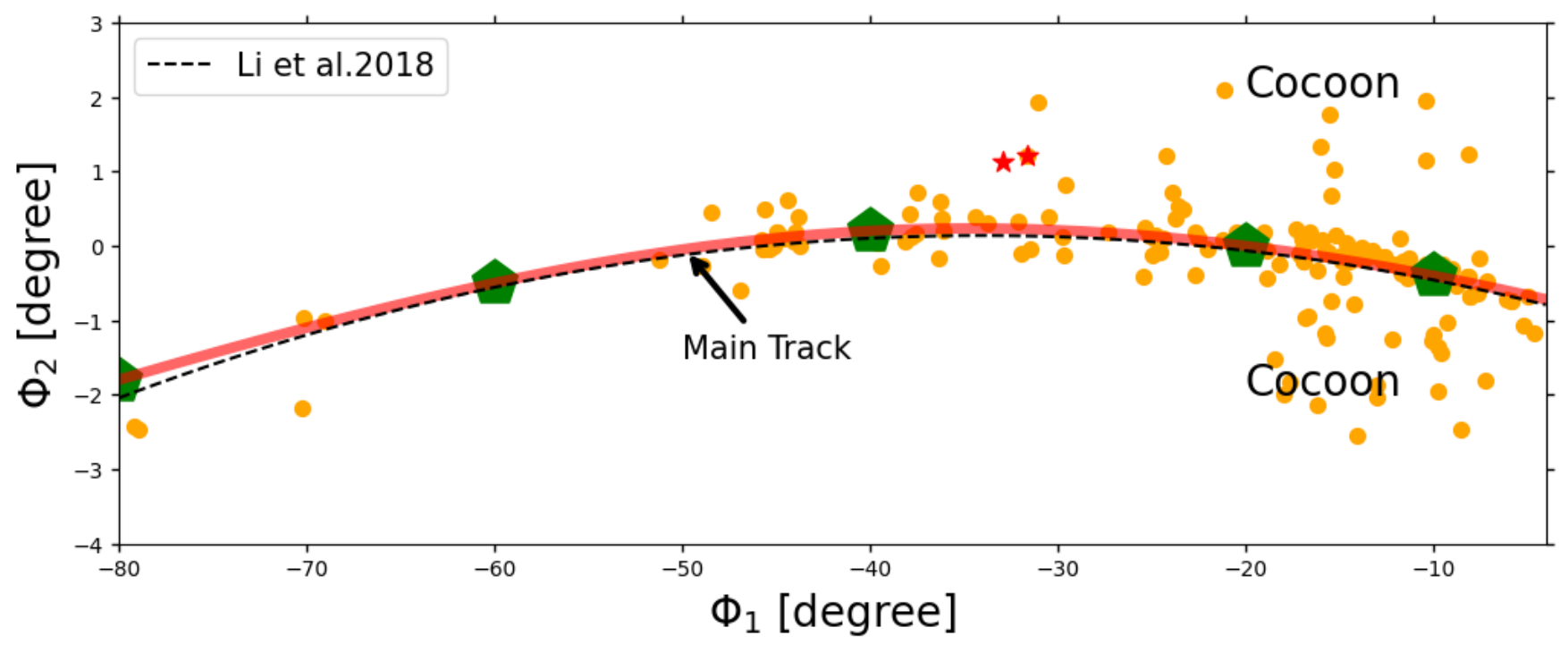}  
        \caption{The distribution of the sample in $\phi_1$–$\phi_2$ coordinates (orange dots), with two spur stars marked as red stars. The red line indicates the main track fitted by a cubic spline with five knots (green pentagons) located at [$-80$,$-1.8$], [$-60$,$-0.5$], [$-40$,$-0.2$], [$-20$,$0.0$], and [$-10$,$-0.4$], adjusted following \citet{Valluri2025}. The black dashed line represents the track obtained by \citet{Li2018} using polynomial fitting for comparison. The diffuse ``cocoon" structure is also labeled in the figure.}
        \label{sample}
    \end{center}
\end{figure*}

We cross-matched the selected member stars with the two catalogs; however, for the sample from \citet{Valluri2025} , we only used abundance estimates from \citet{Zhang2024} for the DESI data. For stars matched in both catalogs, we only consider the DESI abundance estimates rather than the LAMOST ones, in order to maintain consistency in the overall analysis. The LAMOST estimates are used only when DESI abundances are not available. After these selections, the total number of member stars is 176. Among the stars from \citet{Valluri2025}, 66 have DESI estimates. For stars from \citet{Malhan2019GD}, 19 have DESI estimates and 6 have LAMOST estimates. For stars from \citet{Huang2019}, the numbers are 3 and 4, respectively. For stars from \citet{Starkman2025}, there are 73 with DESI estimates and 5 with LAMOST estimates, including two “spur” stars with DESI and LAMOST estimates, respectively. Figure~\ref{sample} shows the distribution of the final sample in GD‑1 coordinates \citep[transformed following][]{Koposov2010}.

\section{Results}\label{results}
Before performing a detailed analysis of GD‑1, we need to separate the two main structures: the cocoon and the thin stream. \citet{Valluri2025} has shown that using a univariate or multivariate Gaussian model does not affect the separation of the cocoon; therefore, similar to \citet{Malhan2019}, we adopt a univariate truncated Gaussian likelihood function, which is:

\begin{equation}
\ln \mathcal{L}(\theta)
=
\sum_{i=1}^{N}
\ln \left[
f_s\,
\frac{1}{\sqrt{2\pi}\,w_s\text{erf}(p,w_s)}
\exp\!\left(
-\frac{(\delta\phi_{2,i} - \mu_s)^2}{2 w_s^2}
\right)
+
f_c\,
\frac{1}{\sqrt{2\pi}\,w_c\text{erf}(p,w_c)}
\exp\!\left(
-\frac{(\delta\phi_{2,i} - \mu_c)^2}{2 w_c^2}
\right)
\right].
\end{equation}

Here $\text{erf}(p,w)$ is the error function and $\delta\phi_2$ is the offset from the main track. $\mu_s$ and $w_s$ are the mean and standard deviation, respectively, of the Gaussian distribution for $\delta \phi_2$ of the stream, while $\mu_c$ and $w_c$ are those for the cocoon. The units for these parameters are degrees. The parameters $f_s$ and $f_c$ represent the fractional contributions (dimensionless) of the stream and cocoon components, respectively. We set $p = 2.35$ based on our sample to truncate $\delta \phi_2$ and assume uninformative priors for all parameters. We used \texttt{emcee} python package \citep{Foreman2013} to sample the posterior distribution of the parameters. During the sampling process, we adopted 100 walkers to explore the five-dimensional parameter space, discarded the first 1000 steps as burn-in, and ran the sampler for an additional 4000 steps for the final parameter estimation. We then evaluated the likelihood using the median values of the model parameters to compute the probability that each star belongs to each component, and selected stars with probabilities greater than 0.8 as members of the corresponding component, which is stricter than the threshold of 0.5 in \citet{Malhan2019}.

\begin{figure*}
    \begin{center}
        \includegraphics[width=1\textwidth]{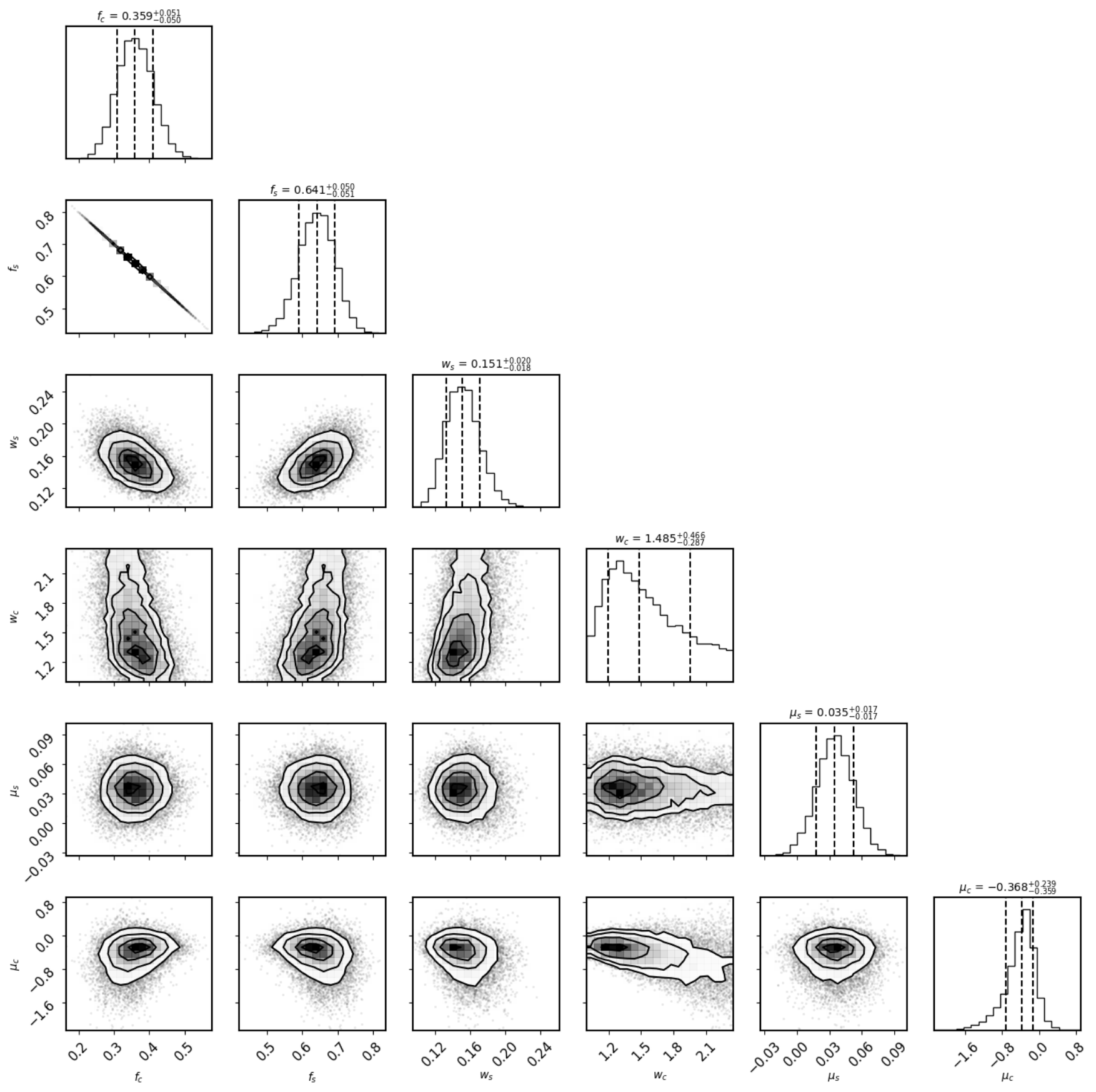}  
        \caption{The posterior distributions of the five parameters. Note that $f_c + f_s = 1$ and the median values of parameters are used to calculate the possibility of membership.}
        \label{posterior}
    \end{center}
\end{figure*}

The posterior distributions of the parameters are shown in Figure~\ref{posterior}. Based on the posterior sampling results, the fraction of the cocoon component is 0.359, while that of the thin stream is 0.641, where the fraction for the thin stream is slightly higher than the fraction in both \citet{Malhan2019} and \citet{Valluri2025}. The higher fraction is likely because our sample contains more member stars and is not subject to the radial velocity selection that would otherwise remove certain stars, which could result in a greater proportion of cocoon stars. However, potential selection effects cannot be ruled out. Based on the results, the thin stream has a full width at half maximum (FWHM) of about 50 pc, while the cocoon has an FWHM of about 488 pc (assuming a mean distance of 8 kpc). This cocoon is wider than that in \citet{Malhan2019}, as \citet{Valluri2025} and the updated literature include a larger number of member stars. In general, our results are consistent with previous studies and are reliable for distinguishing between thin stream and cocoon stars. By selecting memberships with $p>0.8$, there are 46 cocoon stars and 111 stream stars.

\subsection{Chemical Properties of the GD-1 stream}
\subsubsection{No Clear $\alpha$-knee in the GD-1 Stream}
Since the cocoon stars have velocity dispersions similar to the tidal streams of dwarf galaxies \citep{Li2022}, \citet{Valluri2025} also speculated that the cocoon may have been brought in by a massive parent dwarf galaxy through dynamical friction, while the thin stream was from a nuclear cluster or a GC. Because Type Ia supernovae have a longer time-scale than Type II supernovae, leading to delayed contributions, they typically produce an ``$\alpha$-knee" in the [$\alpha$/Fe]–[Fe/H] plane during galactic evolution \citep[e.g.,][]{Mason2024}. Traces of this $\alpha$‑knee can also be found in phase-mixed substructures \citep{Monty2020,Horta2023}. Therefore, one of the best ways to test this hypothesis is to study the evolution of its $\alpha$‑element abundances (e.g., [Mg/Fe] as a proxy).

\begin{figure*}
    \begin{center}
        \includegraphics[width=1\textwidth]{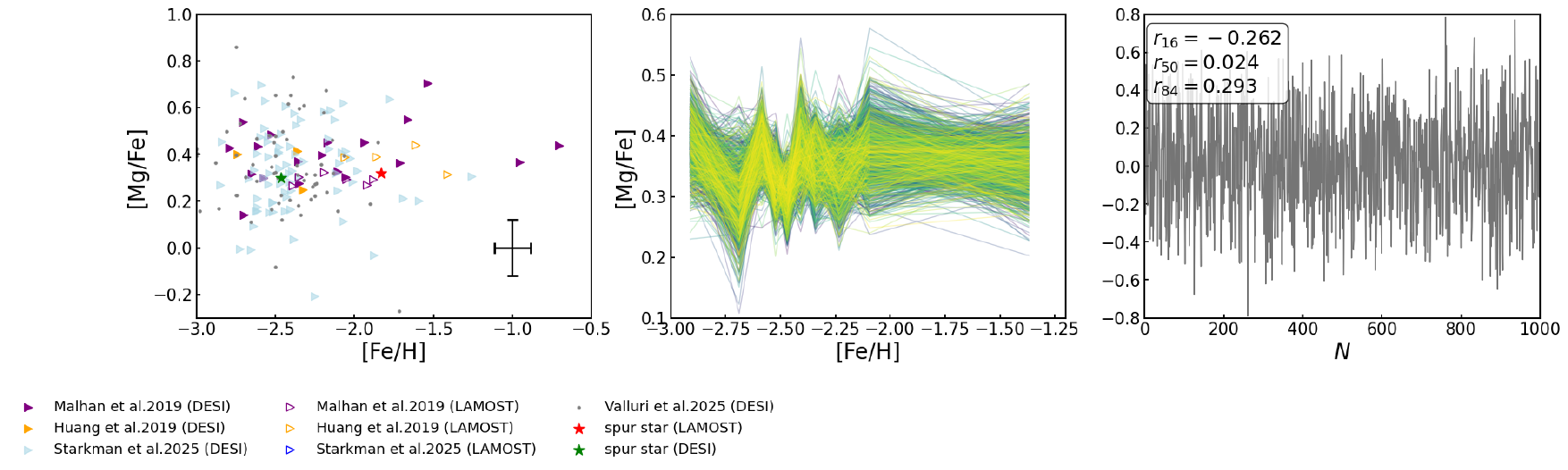}  
        \caption{Left panel: the [Mg/Fe]–[Fe/H] distribution of the GD-1 stream, with Mg abundances cross-matched from \citet{Zhang2024} and \citet{Liu2025}. The black error bar represents the median uncertainties of the sample.  Middle panel: 1000 Monte Carlo realizations of the median [Mg/Fe] trends. Right panel: the Pearson correlation coefficient for each realization, with the 16th, 50th, and 84th percentiles of the coefficients indicated.}
        \label{mgfe}
    \end{center}
\end{figure*}

The left panel of Figure~\ref{mgfe} shows the [Mg/Fe]–[Fe/H] distribution of GD-1 member stars after cross-matching the two catalogs. No calibration was applied to the abundances, as the DESI measurements dominate (166 stars). The mean [Mg/Fe] of the member stars is $\sim$ 0.35, slightly higher than the 0.3 reported by \citet{Balbinot}. To account for the uncertainties in [Mg/Fe], we performed Monte Carlo sampling of each star’s [Mg/Fe] values assuming a normal distribution, and analyzed the resulting median [Mg/Fe] trend over 1000 realizations (middle panel). We still find that these trends show no obvious downturn after the plateau. The right panel presents the variation of the Pearson correlation coefficient between the median [Mg/Fe] trend and [Fe/H] across the 1000 realizations. It can be seen that the coefficient remains positive for most realizations, indicating that the GD-1 stream is indeed likely to lack a clear $\alpha$-knee. Therefore, these results argue against this hypothesis, since more massive dwarf galaxies typically experience significant star formation and thus exhibit an $\alpha$-knee.

However, one might argue that the parent dwarf galaxy of GD-1 could have been accreted very early, so that Type Ia supernovae had not yet contributed significantly. As is commonly considered for the Gaia–Sausage–Enceladus substructure \citep{Berokurov2018,Helmi2018}, which was accreted during a major merger event early on (estimated at $8-11$ Gyr ago), it also exhibits an $\alpha$-knee \citep{Horta2023} and is largely phase-mixed. Therefore, such a scenario would have difficulty explaining why GD-1 still maintains a coherent stellar stream structure. Furthermore, GD-1 exhibits a very narrow [Fe/H] distribution, with a dispersion of $\sim$ 0.10 dex \citep{Shi2025}, which is also typical of GCs. Therefore, the formation of a cocoon via the dynamical friction of a dwarf galaxy is unlikely, and this scenario would be best tested with dedicated cosmological simulations

\begin{figure*}
    \begin{center}
        \includegraphics[width=1\textwidth]{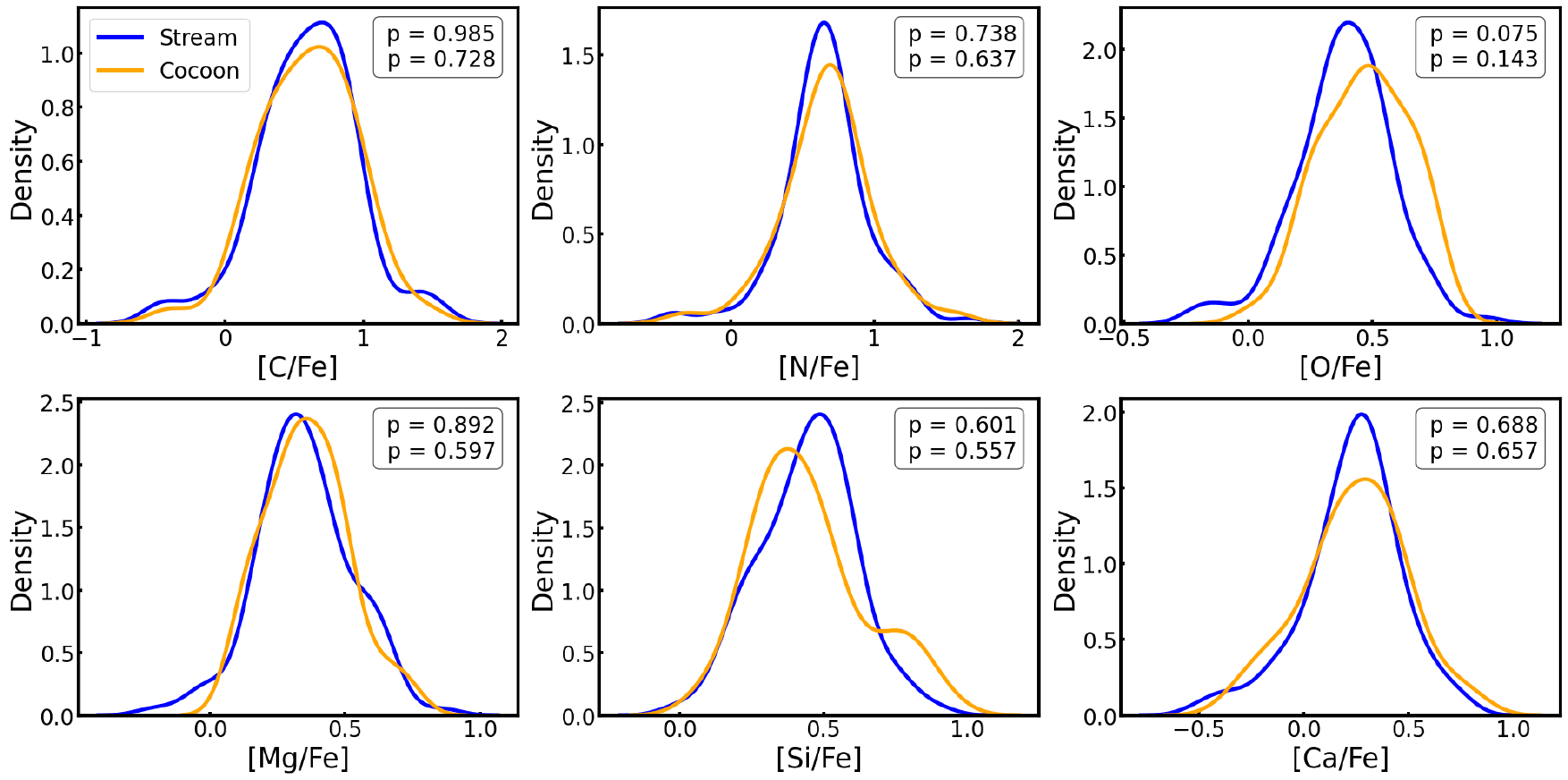}  
        \caption{The KDE plots for the stream (blue) and cocoon (orange) stars among six elements. The p-values in the first row are the KS test results without uncertainties, and the p-values in the second row are the results considering Gaussian perturbations with an uncertainty of 0.2 dex.}
        \label{element}
    \end{center}
\end{figure*}

\subsubsection{The Common Origin of the Cocoon and Thin Stream}
Previous studies using simulations have shown that the cocoon may originate from pre-stripping within the parent dark matter sub-halo \citep[e.g.,][]{Malhan2019,Qian2022}. The similarity in [Fe/H] between the cocoon and the thin stream further supports a common origin for both components. Here, we further confirm their common origin by incorporating additional elemental abundances. Figure~\ref{element} shows the kernel-density estimation (KDE) distributions of six elements for 46 cocoon stars and 111 stream stars. Among these, only one cocoon member star has LAMOST elemental abundance estimates, whereas 11 stream member stars have elemental abundance estimates from LAMOST. It can be well observed that cocoon and stream have similar distributions across all elements, and according to the Kolmogorov–Smirnov (KS) test, their p-values do not reject the hypothesis that the two come from the same distribution (p-values in the first row). Moreover, since the elemental abundances inferred by deep learning often underestimate the true uncertainties, we added Gaussian perturbations with an uncertainty of 0.2 dex to these elements. This 0.2 dex uncertainty is slightly larger than the median uncertainties of most elements provided by the two matched catalogs. It can be seen that, still under the KS test, the p-values do not reject the null hypothesis (p-values in the second row of Figure~\ref{element}), further confirming the common origin of cocoon and stream.

\subsubsection{No Clear MP Signatures and Comparisons with Extragalactic GCs }
In most observed GCs, two distinct stellar populations are present: first-population (1P) stars, which resemble field stars, and second-population (2P) stars, which show signatures of high-temperature H-burning \citep{Milone2022}. Although stars formed within a given GC are generally expected to share similar chemical abundances due to their common birth environment, significant spreads in light-element abundances and He variations are observed, indicating distinct formation and evolutionary pathways for 1P and 2P stars. The unique probes for MP signatures are Na-O, C-N and Mg-Al anti-correlations as well as Na–N and C-O correlations, which are useful for studying GCs. On the other hand, low-mass GCs ($\sim10^4 M_\odot$) may not exhibit clear MP signatures; however, the small number of such clusters studied to date prevents any firm conclusion, with the exception of GC E3 \citep{Salinas2015}.

The initial mass of the GD-1 stream is assumed to be $\sim1.58\times10^4M_\odot$ \citep{deboer2020}, which indicates a low-mass GC progenitor for the GD-1 stream. Therefore, a study of the elemental abundances in GD-1 can further confirm and constrain the mass of its progenitor. In a recent study, \citet{Zhao2025} found no discernible Na-O anti-correlation in the member stars, but the sample was very limited and only two stars had available O measurements. Although elemental abundances predicted by neural networks may have relatively large uncertainties, they provide a much larger sample base and thus increase statistical significance. In terms of applications to GCs, for example, \citet{Kane2025} employed a neural network to identify stars with a GC origin.

\begin{figure*}
    \begin{center}
        \includegraphics[width=1\textwidth]{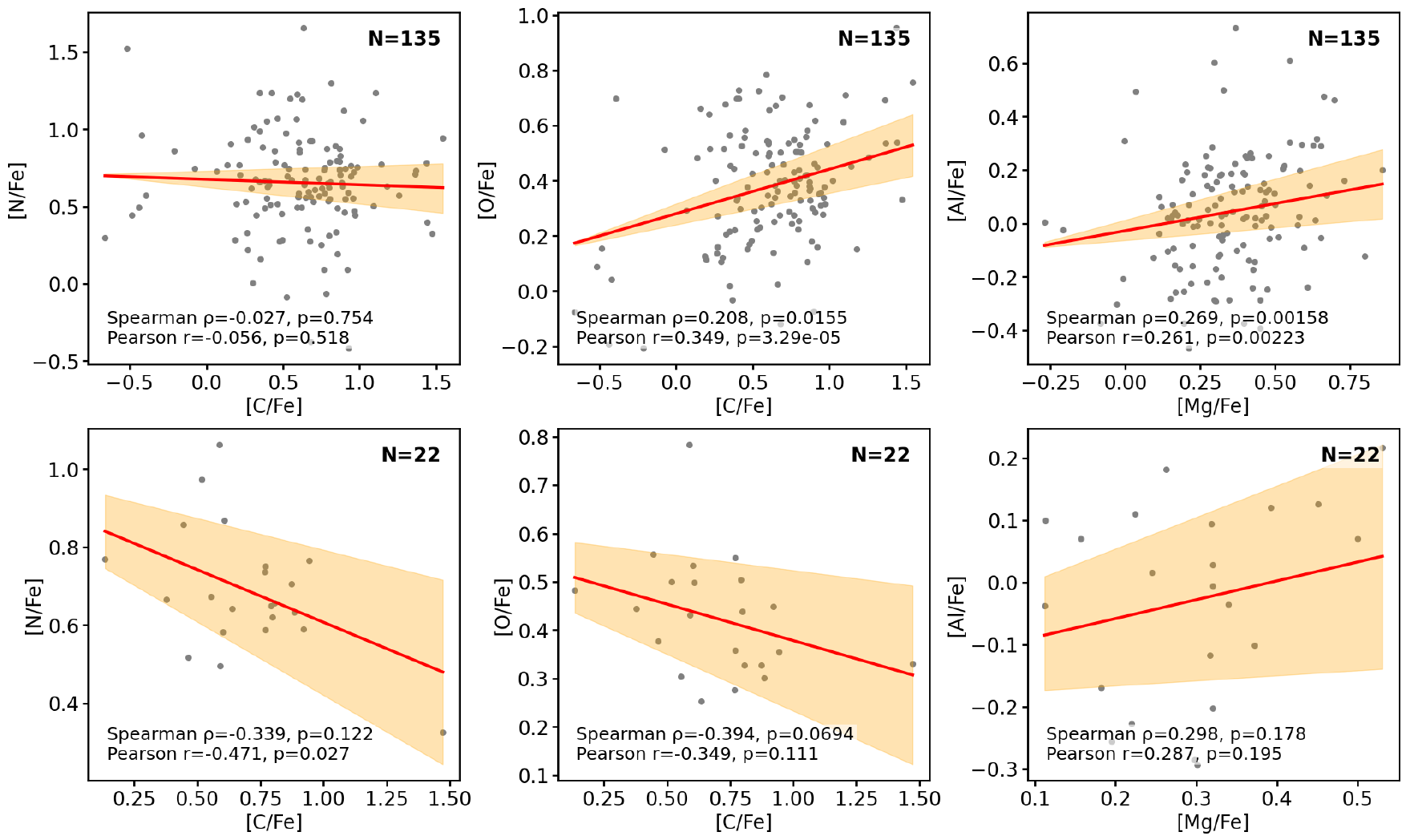}  
        \caption{The distributions of GD-1 member stars, considering only the abundances from \citet{Zhang2024}, are shown for the C-N, C-O, and Mg-Al relations. The red line represents a linear fit obtained from 1,000 bootstrap re-samplings, and the orange shaded area corresponds to the 16th and 84th percentiles of the distribution. Both Spearman and Pearson correlation coefficients, along with their p-values, are indicated, as well as the sample size. The samples in the second row are further restricted by an error cut requiring errors to be $<0.15$ dex.}
        \label{MP}
    \end{center}
\end{figure*}

\begin{figure*}
    \begin{center}
        \includegraphics[width=1\textwidth]{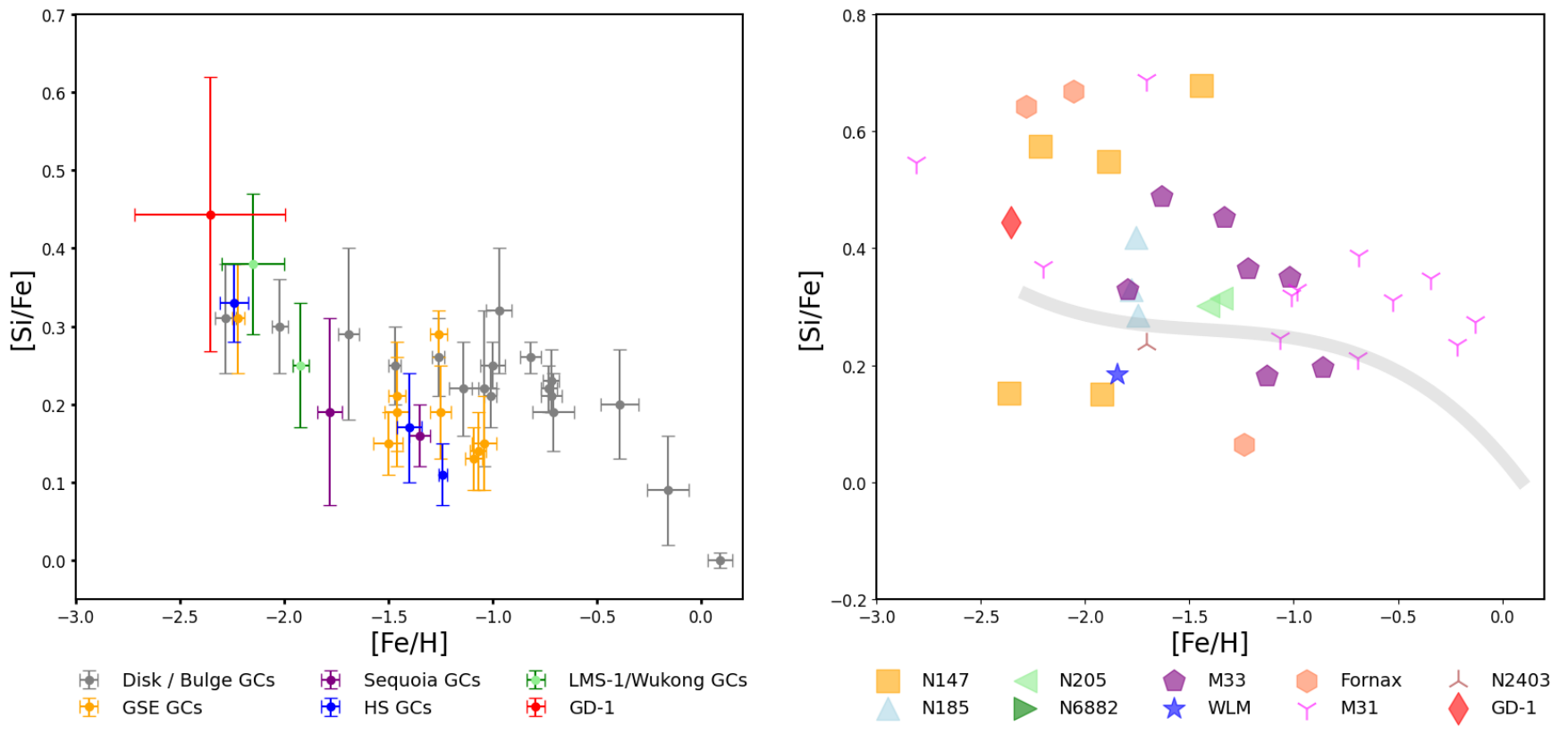}  
        \caption{Left panel: the distributions of the GD-1 and GCs related to known substructures in the [Fe/H]-[Si/Fe] plane, where the error bars indicate one standard deviation. Right panel: similar to left panel (but for GCs in the Local Group), the gray line shows the distribution range of the MW GCs. }
        \label{extra}
    \end{center}
\end{figure*}

To minimize selection bias and systematic errors, we consider only the sample (135 stars) drawn from \citet{Valluri2025} and adopt elemental abundances estimated exclusively by \citet{Zhang2024}. As shown in Figure~\ref{MP}, when uncertainties are not considered (the first row), C-N shows an extremely weak and non-significant correlation in terms of both Spearman and Pearson coefficients, C-O exhibits a slightly significant positive correlation, while Mg-Al shows a positive correlation opposite to that of the MP signature. After applying a cut with errors $<$ 0.15 dex (the second row), although C-N begins to show a negative correlation, the relationships for C-O and Mg-Al still contradict the MP signature, and the significantly reduced sample size limits the statistical analysis. However, these trends still suggest that the GD-1 stream is very likely devoid of MP signatures and originates from a low-mass progenitor.

Previous studies have also shown that the GD-1-like streams have an ex-situ origin \citep{Malhan2019,Qian2022,Valluri2025}. To investigate the possibility that the GD-1 progenitor was accreted, we compare the GD-1 stream with ex-situ GCs related to known substructures as well as those in the Local Group in the [Fe/H]-[Si/Fe] plane \citep{Horta2020,Larsen2022}. As seen from the left panel of Figure~\ref{extra}, GD-1 has relatively large error bars due to the high uncertainties in the deep-learning methods. The very low metallicity of GD‑1, combined with its relatively high [Si/Fe] abundance, further supports its old age \citep[$\sim13$ Gyr,][]{Li2018}. Moreover, GD-1 does not exhibit abundance patterns similar to those of GCs known to originate from disrupted dwarf galaxies, and its distribution is entirely distinct from that of the MW GCs, suggesting that it originated from an as-yet unidentified dwarf galaxy. As can also be seen from the right panel, GD‑1 exhibits a distribution similar to that of GC residing in dwarf galaxies, indicating that GD‑1 is very likely to have an ex-situ origin. We also note that the two GCs of N185 exhibit similarly low metallicities and $\alpha$-element abundances higher than those of the Milky Way. If the mass of the GD-1 parent dwarf galaxy was comparable to that of N185 \citep[$\sim10^8M_\odot$,][]{Hamedani2017} and it was accreted long ago (consistent with the old ages of the GD-1 stars), it is therefore not surprising that no substructures associated with GD‑1 have been found, as these debris from such a small galaxy may have already been fully phase-mixed. In summary, we conclude that GD‑1 is of ex-situ origin and that its progenitor dwarf galaxy was likely of low mass and accreted at an early time.

\subsection{Sagittarius-induced Dynamical Perturbations of the GD-1 Stream}
\subsubsection{The Setup of Test-particle Simulations }
Both \citet{Bonaca2020} and \citet{Dillamore2022} have shown that the Sgr has a great impact on the morphology of GD-1-like streams and could be responsible for creating off-track features. Based on this scenario, we plan to carry out test-particle simulations and, using the particle-spray algorithm developed by \citet{Chen2025} implemented in \texttt{galpy} python package \citep{Bovy2015}, investigate whether the Sgr could have formed a cocoon-like structure.
\begin{figure*}
    \begin{center}
        \includegraphics[width=0.9\textwidth]{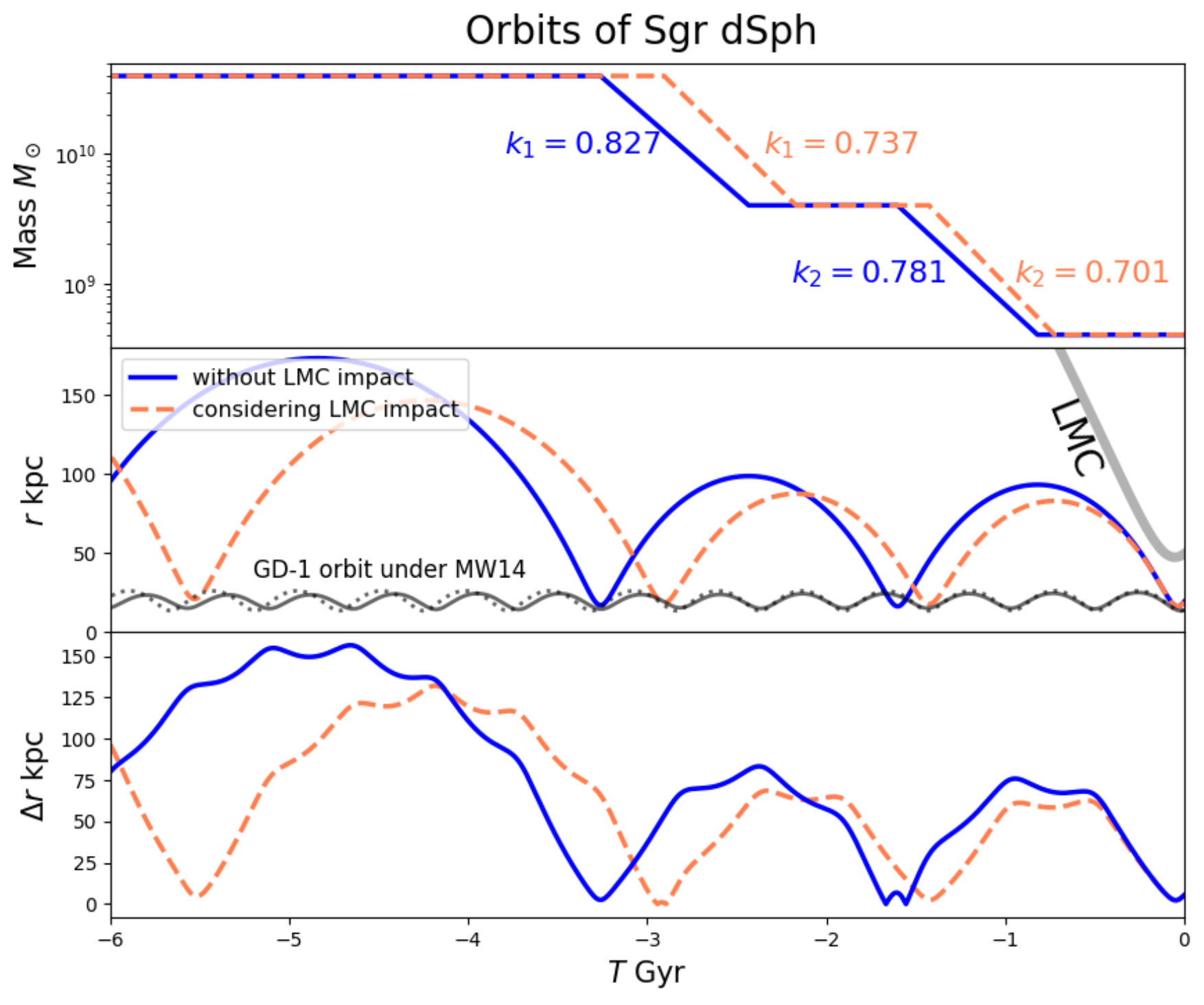}  
        \caption{Upper panel: the mass variations of the Sgr with the LMC (orange) and without the LMC (blue), where $k$ indicates the slope during the mass-loss period for the corresponding color. Middle panel: the orbits of the Sgr with the LMC (orange) and without the LMC (blue), and the orbits of the GD-1 progenitor with the Sgr (black line) and without the Sgr (black dotted line). Lower panel: the distances between the Sgr and the GD-1 progenitor with the LMC (orange) and without the LMC (blue). }
        \label{setup}
    \end{center}
\end{figure*}

\begin{figure*}
    \begin{center}
        \includegraphics[width=1\textwidth]{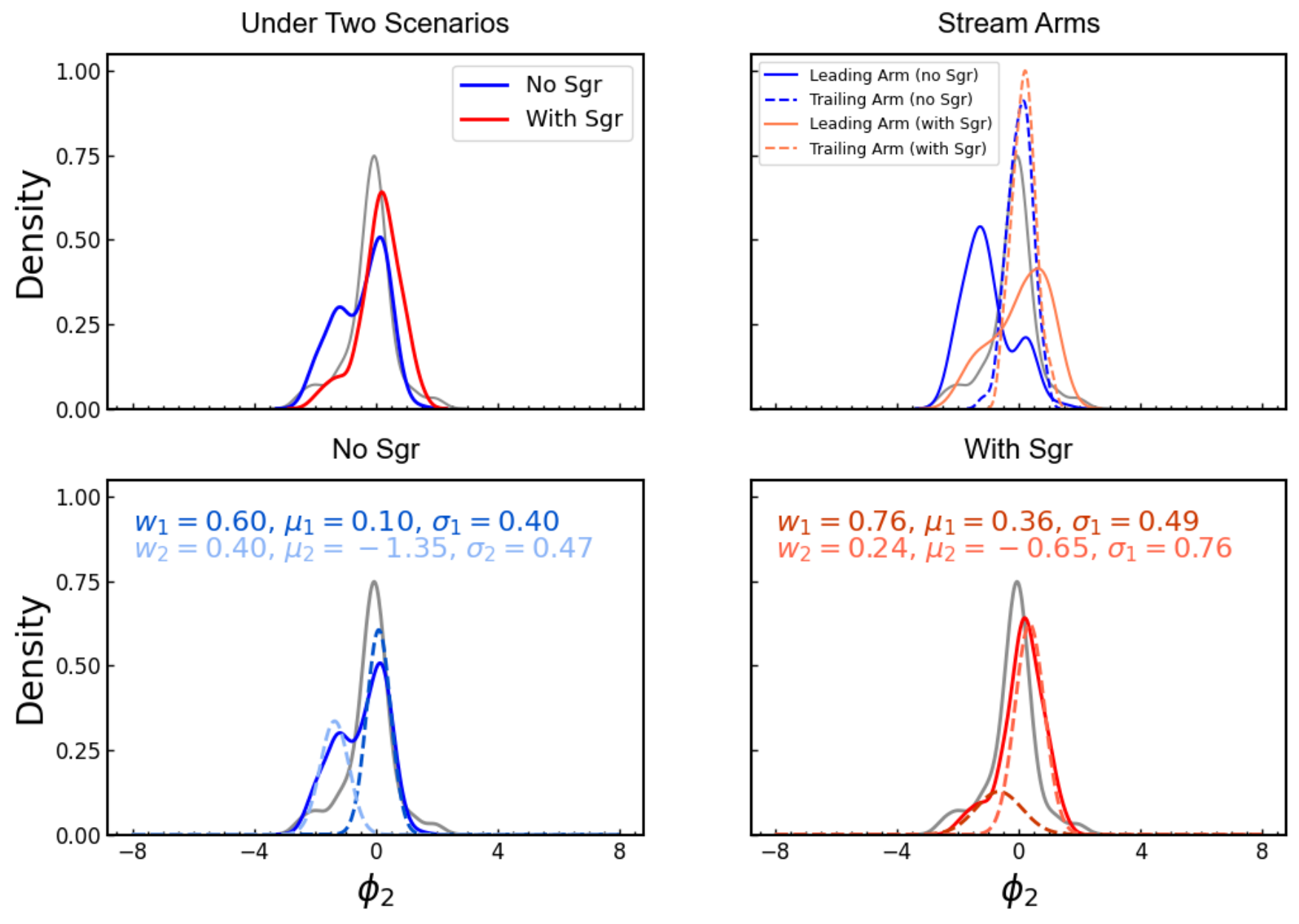}  
        \caption{Top-left panel: the density distributions along $\phi_2$ coordinate with the Sgr (red) and without the Sgr (blue). Top-right panel: the density distributions along $\phi_2$ coordinate for the leading (solid lines) and trailing arms (dashed lines) under two models. Bottom row: the two Gaussian components of the streams with and without the Sgr, with the weights, means, and standard deviations of each component indicated. The gray solid line represent the distributions of the observational data.}
        \label{density}
    \end{center}
\end{figure*}
In the simulation, we require the mass of the Sgr to be time-dependent. That is, during the passage from pericenter to apocenter, its logarithmic mass decays with a decay factor $\delta$, whereas between apocenter and the next pericenter, the mass remains constant. Here, we directly choose $\delta$ = 1, because \citet{Dillamore2022} found that in this case the likelihood of matching the observational data is maximized. Similarly, following \citet{Dillamore2022} we adopt the phase-space coordinates and current mass ($4\times10^8M_\odot$) of the Sgr used by them, and assume a Hernquist profile for the Sgr with a varying scale radius $a\propto M^{1/3}$ (current value 1.4 kpc). As for the dynamical friction \citep{Chand1943} of the MW and the Sgr, we also use the same velocity dispersion $\sigma=120$ km/s and Coulomb logarithm, employing \texttt{ChandrasekharDynamicalFrictionForce} function (\texttt{cdf} for short) implemented in \texttt{galpy}. For the Large Magellanic Cloud (LMC), we also adopt a Hernquist profile and assume a mass of $1.5\times10^{11}M_\odot$, which is within the range given by \citet{Shipp2021}, and a half-mass radius of 5 kpc. For the GD-1 progenitor, we adopt a Plummer profile with a mass of $2\times10^4M_\odot$ and a scale radius of 2 pc \citep[consistent with][]{Dillamore2022} and phase-space coordinates from \citet{Webb2019}. Although GD-1 shows no clear MP signature, indicating that its progenitor GC had a very small mass, we find that changing the mass mainly affects the physical length of the generated stream without having a larger impact, and therefore we do not adjust its mass. For the MW potential, we choose \texttt{MilkyWayPotential2014} \citep[\texttt{MW14} for short;][]{Bovy2015} as this potential best matches the observed acceleration field \citep{Nibauer2025}.

During the testing phase, we found that although the LMC does not affect the orbit of GD-1, it significantly influences the orbit of the Sgr, and therefore its effect must be taken into account (see Figure~\ref{setup}). During the orbital integration, the time-dependent potentials of moving galaxies are characterized by \texttt{MovingObjectPotential} function. For the LMC, its orbit is integrated in \texttt{MW14} + \texttt{cdf}. For the Sgr, its orbit is integrated in \texttt{MW14} + \texttt{cdf} + \texttt{NIP} (non-inertial potential) + \texttt{moving\_LMC}. The \texttt{NIP} is included because the large mass of the LMC can significantly shift the center of mass of the MW. Then the GD-1 progenitor is integrated in \texttt{MW14} + \texttt{moving\_LMC} + \texttt{moving\_Sgr}, while the generated stream particles are integrated in \texttt{MW14} + \texttt{moving\_LMC} + \texttt{moving\_Sgr} + \texttt{moving\_GD1}. They are all integrated backward 6 Gyr.

Figure~\ref{setup} shows the orbits of the Sgr and GD-1. As mentioned above, the LMC strongly affects the orbit of the Sgr by shortening its orbital period and reducing its apocentric distance, whereas the orbit of GD-1 is not significantly influenced by the gravitational potential of the Sgr, showing only a noticeable phase shift at early times (the middle panel). We also note that the Sgr experiences a pericentric passage at around 5.5 Gyr ago, whereas in the model of \citet{Dillamore2022} this passage occurs near 6 Gyr, which may be due to differences in the adopted gravitational potential models. In addition, we no longer consider mass variations of the Sgr prior to 5.5 Gyr ago, since its influence during this period is outside the scope of our analysis, given that the dynamical age of the GD-1 stream is only about $2-3$ Gyr \citep{Webb2019}. Interestingly, we note that at $\sim$ 3 Gyr and 1.4 Gyr ago, the Sgr and the GD-1 progenitor experienced close encounters (considering the LMC potential), which may have had some impact on the GD-1 stream. 

Given the dynamical age of the system and the limitations of the method itself, we perform particle spray using the \texttt{chen24spraydf} function \citep{Chen2025} within a time window of ($-2.009$, $-1.999$) Gyr, corresponding to $\Delta$t = 10 Myr. We sample 100 particles for each of the leading and trailing arms. During this window, the mass of the Sgr is kept fixed at $4\times10^9 M_\odot$. Although individual particles are stripped at slightly different times, the narrow time window allows us to approximate that all particles are stripped at $-1.999$ Gyr and are then integrated forward in time. As a result, the Sgr may have influenced the particles prior to $\sim$ 1.4 Gyr ago. For comparison, we also carry out simulations that do not include the Sgr and the results are shown below.

\subsubsection{The Results of Test-particle Simulations}
\begin{figure*}
    \begin{center}
        \includegraphics[width=1\textwidth]{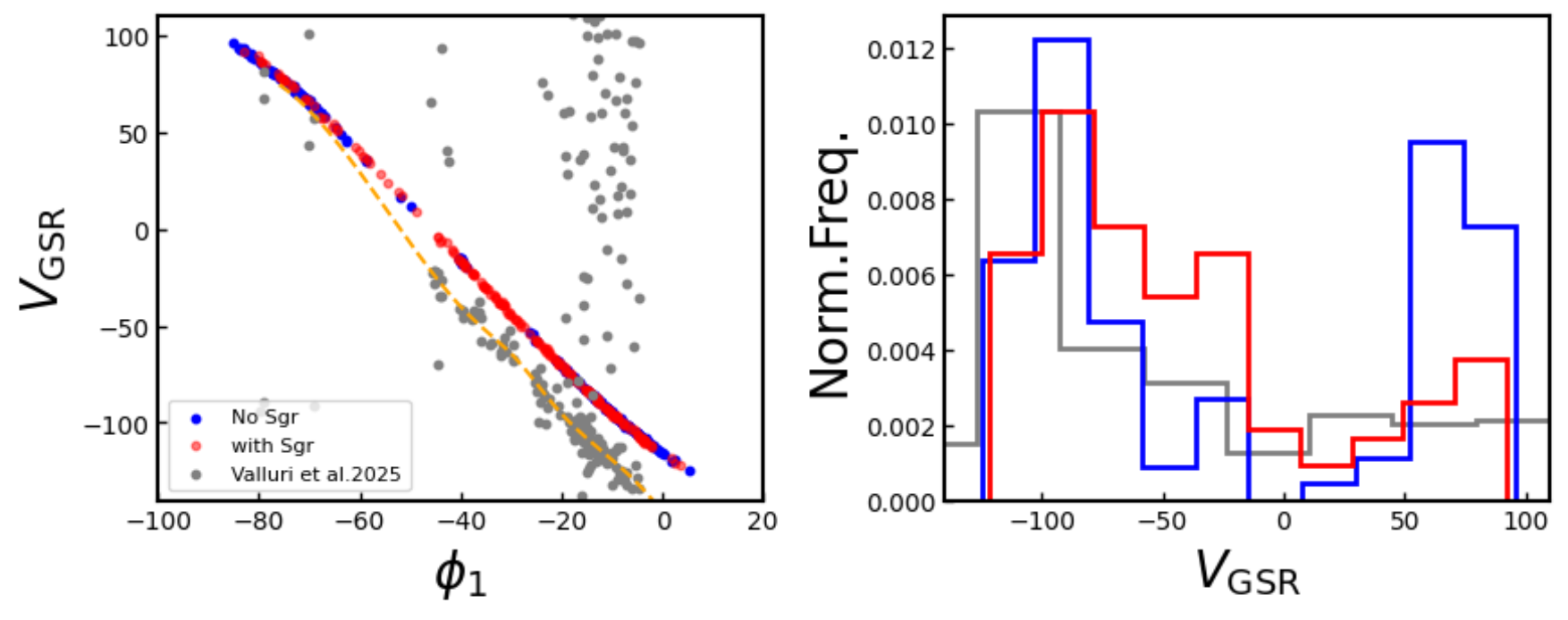}  
        \caption{Left panel: the distributions of stream particles in the $V_{\text{GSR}}$-$\phi_1$ space with (red) and without the Sgr (blue), while the gray dots and orange dashed line are the data and spline from \citet{Valluri2025}. Right panel: the histograms of $V_{\text{GSR}}$, where the colors stand for the same meanings.}
        \label{GSR}
    \end{center}
\end{figure*}

Figure~\ref{density} and Figure~\ref{GSR} illustrate the results of our simulations. The first row of Figure~\ref{density} shows the density distribution of stream particles in $\phi_2$ with and without the Sgr, as well as the $\phi_2$ distributions of the leading and trailing arms. It can be indicated that encounters with the Sgr do not significantly heat the stream particles, so as to broaden their $\phi_2$ distribution. When the Sgr is included, the simulated $\phi_2$ distribution matches the GD-1 observational data much better. In contrast, in the model without the Sgr, an additional smaller secondary bump appears ($\phi_2\sim-1.5^{\circ}$), which is not clearly seen in the observations.
Moreover, the thin component of GD-1 closely follows the $\phi_2$ distribution of the trailing arm in both models, suggesting that the main contribution to the thin stream likely comes from the trailing arm. In contrast, the leading arm consistently exhibits a broader distribution in $\phi_2$; in the model with the Sgr, its distribution is more symmetric relative to the thin stream, whereas in the model without the Sgr, it is overall shifted toward smaller $\phi_2$, and this shift is responsible for the aforementioned secondary bump.

We also find that, according to the Bayesian Information Criterion \citep[BIC;][]{BIC}, the streams produced by both models are best described as being composed of two Gaussian components. However, in the presence of the Sgr, the distributions of the Gaussian components show better quantitative agreement with GD-1, but the fraction of the cocoon component remains relatively low ($\sim$24 compared to $\sim$36 in the data), indicating that other heating mechanisms may still be at play. According to the model without the Sgr, the dispersion in the $\phi_2$ direction may primarily arise from the intrinsic dynamics of the stream itself, such as the stripping initial conditions and long-term orbital evolution. However, perturbations from the Sgr have a non-negligible effect on reshaping this distribution. Taking these results into account, the Sgr is more likely to influence the spatial and statistical distribution of the ``diffuse component” in GD-1 rather than its formation mechanism.

We also examined the effect of the Sgr on the distribution of velocities in the Galactic standard of rest ($V_{\text{GSR}}$)\footnote{calculations are based on https://docs.astropy.org/en/latest/coordinates/example\_gallery\_rv\_to\_gsr.html}. We can see that the presence or absence of the Sgr does not affect the distribution of stream particles in the $V_{\text{GSR}}$-$\phi_1$ space (left panel of Figure~\ref{GSR}). This may also be because we only consider particles stripped within a relatively short time window, so there is no significant dispersion in this space. Compared to the data from \citet{Valluri2025}, our stream particles have higher velocities at the same $\phi_1$. Although the Sgr has little effect in the $V_{\rm GSR}$–$\phi_1$ space overall, it pulls stream particles toward the Galactic center, causing more particles to accumulate in the $V_{\rm GSR} < 0$ range and improving the agreement with observations (right panel). In the absence of the Sgr, a prominent secondary feature appears near $V_{\rm GSR} \sim 75$ km/s. Even in the presence of the Sgr, there is a less pronounced peak at this location (compared to the observations), indicating that other mechanisms also influence this part of the stream.

Therefore, considering only the effect of the Sgr can already explain some features of GD-1 reasonably well. However, in our simulations, we only spray particles within a very short time window. In reality, the size of this window may also affect the results. A more realistic treatment would also need to account for interactions with dark matter sub-halos, which will be addressed in future work.

\section{Summary}\label{summary}
In this work, we use elemental abundance estimates obtained from deep learning to study the chemical properties of the GD-1 stream member stars over a relatively large sample, comprising up to 176 stars. After 1000 Monte Carlo realizations, GD-1 exhibits a roughly horizontal trend in the [Mg/Fe]-[Fe/H] distribution, with no clear $\alpha$-knee. Moreover, the corresponding [Mg/Fe] median trends show positive Spearman coefficients in most cases. This indicates that GD-1 is unlikely to have originated from a massive galaxy undergoing significant dynamical friction and instead points to a clear GC origin. In addition, the separated cocoon and thin stream stars exhibit similar chemical abundance distributions, further confirming a common origin. At the same time, we also find that the GD-1 stream shows no obvious MP signatures, indicating that its progenitor had a very low mass; however, this requires further star-to-star spectroscopic confirmation of elemental abundances. Furthermore, the abundances of GD-1 are similar to those of GCs in the Local Group, suggesting that it was likely accreted. 

Finally, test-particle simulations show that the influence of the Sgr is non-negligible. Although the Sgr does not significantly heat the stream to form a ``cocoon" structure, it has a substantial effect on the spatial distribution of the stream along $\phi_2$ coordinate. In addition, the thin stream is likely mainly contributed by the trailing arm, while the cocoon component comes from the leading arm. Moreover, the Sgr does not affect the distribution of particles in the $\phi_1$–$V_{\rm GSR}$ space in the simulations, but it pulls stream particles toward the Galactic center, causing more particles to accumulate in the $V_{\rm GSR} < 0$ range and improving agreement with observations. However, it is worth noting that our simulation only considered a single time window and did not explore the effects of varying stripping durations. Additionally, the impacts of dark matter sub-halos and the pre-stripping of the globular cluster within its parent halo cannot be overlooked. Therefore, future work will need to provide a more comprehensive consideration of these factors. In all, the results of our work provide new constraints on the GD-1 stream both chemically and dynamically.

\section{ACKNOWLEDGMENTS}
We thanks Bill Chen for the useful suggestions of using the particle-spray method. We also thanks the insightful comments of the reviewer that have greatly improved this manuscript. This work was supported by National Key R\&D Program of China No. 2024YFA1611900, and the National Natural Science Foundation of China (NSFC Nos. 11973042).

Software: \texttt{NumPy}\citep{Harris2020}, \texttt{Matplotlib}\citep{Hunter2007}, \texttt{Astropy}\citep{astropy2013,astropy2018,astropy2022}, \texttt{galpy} \citep{Bovy2015}.

\bibliography{liu}{}
\bibliographystyle{aasjournal}



\end{document}